\documentclass{article}
\usepackage{amsmath}
\usepackage{amssymb}
\usepackage{amsfonts}
\usepackage[backend=bibtex,
  style=ieee,
  sorting=none,
  url=false,
  style=numeric,
  maxbibnames=99,
firstinits=true]{biblatex}
\renewbibmacro{in:}{}
\bibliography{/home/rsrsl/library/jabref_biblio.bib}
\usepackage{enumerate}
\usepackage{graphicx}
\graphicspath{{figures/}}
\usepackage{float}
\usepackage{algorithm}
\usepackage{xcolor}
\newcommand{\norm}[1]{\left\| #1 \right\|}

\newcommand{\E}{\mathbb{E}}
\newcommand{\N}{\mathcal{N}}

\newcommand{\trace}{\mathrm{Tr}}
\newcommand{\diag}{\mathrm{diag}}
\newcommand{\T}{\mathbf{T}}
\newcommand{\R}{\mathbf{R}}
\newcommand{\kron}{\otimes}
\newcommand{\vect}{\mathrm{vec}}
\renewcommand{\matrix}[1]{\begin{bmatrix} #1 \end{bmatrix}}

\newcommand{\var}{\mathrm{var}}

\newtheorem{theorem}{{\bf Theorem}}
\newtheorem{rem}{{\bf Remark}}

\title{\LARGE \bf Blind system identification\\using kernel-based methods$^\star$} 
\author{Giulio Bottegal, Riccardo S. Risuleo, and H\r akan Hjalmarsson
\thanks{G. Bottegal, R. S. Risuleo and H. Hjalmarsson are with the ACCESS
Linnaeus Center, School of Electrical Engineering, KTH Royal Institute of
Technology, Sweden ({\tt \small risuleo;bottegal;hjalmars@kth.se}).  This work
was supported by the European Research Council under the advanced grant LEARN,
contract 267381 and by the Swedish Research Council under contract
621--2009--4017.
}}
\begin{document}
\maketitle


\begin{abstract}                          
We propose a new method for blind system identification (BSI). Resorting to a Gaussian regression framework, we model the impulse response of the unknown linear system as a realization of a Gaussian process. The structure of the covariance matrix (or kernel) of such a process is given by the stable spline kernel, which has been recently introduced for system identification purposes and depends on an unknown hyperparameter. We assume that the input can be linearly described by few parameters. We estimate these parameters, together with the kernel hyperparameter and the noise variance, using an empirical Bayes approach. The related optimization problem is efficiently solved with a novel iterative scheme based on the Expectation-Maximization (EM) method. In particular, we show that each iteration consists of a set of simple update rules. We show, through some numerical experiments, very promising performance of the proposed method.
\end{abstract}

\section{Introduction}
In many engineering problems where data-driven modeling of dynamical systems is required, the experimenter may not have access to the input data. 
In these cases, standard system identification tools such as PEM~\cite{ljung1999system} cannot be applied and specific methods, namely
\emph{blind system identification} (BSI) methods (or \emph{blind deconvolution},
if one is mainly interested in the input), need be employed~\cite{abedmeraim1997blind}. 

BSI finds applications in a wide range of engineering areas, such as image
reconstruction~\cite{ayers1988iterative}, biomedical sciences
~\cite{mccombie2005laguerre} and in particular communications~\cite{gustafsson1995blind,moulines1995subspace}, for which literally hundreds of methods have been developed. It would be impossible to give a thorough literature review here.

Clearly, the unavailability of the input signal makes BSI problems generally
ill-posed. Without further information on the input sequence or the structure of
the system, it is impossible to retrieve a unique description of the system~\cite{tong1991indeterminacy}. To circumvent (at least partially) this intrinsic
non-uniqueness issue, we shall assume some prior knowledge on the input.
Following the framework of~\cite{ohlsson2014blind} and~\cite{ahmed2014blind}, we
describe the input sequence using a number of parameters considerably smaller
than the length of the input sequence; see Section~\ref{sec:BSI} for details and applications.

The main contribution of this paper is to propose a new BSI method. Our system
modeling approach relies upon the kernel-based methods for linear system
identification recently introduced in a series of papers~\cite{pillonetto2010new,pillonetto2011prediction,chen2012estimation,pillonetto2014kernel}. The main advantage of these
methods, compared to standard parametric methods~\cite{ljung1999system}, is that the user
is not required to select the model structure and order of the system, an
operation that might be difficult if little is known about the dynamics of the
system. Thus, we model the impulse response of the unknown system as a
realization of a Gaussian random process, whose covariance matrix, or
\emph{kernel}, is given by the so called \emph{stable spline kernel}~\cite{pillonetto2010new,bottegal2013regularized}, which encodes prior information on BIBO
stability and smoothness. Such a kernel depends on a \emph{hyperparameter} which
regulates the exponential decay of the generated impulse responses. 

In the kernel-based framework, the estimate of the impulse response can be
obtained as its Bayes estimate given the output data. However, when applied to
BSI problems, such an estimator is a function of the kernel hyperparameter, the
parameters characterizing the input and the noise variance. All these parameters
need to be estimated from data. In this paper, using empirical Bayes arguments,
we estimate such parameters by maximizing the marginal likelihood of the output
measurements, obtained by integrating out the dependence on the system. In order
to solve the related optimization problem, which is highly non-convex, involving
a large number of variables, we propose a novel iterative solution scheme based
on the Expectation-Maximization method~\cite{dempster1977maximum}. We show that each
iteration of such a scheme consists of a sequence of simple updates which can be
performed using little computational efforts. Notably, our method is completely
automatic, since the user is not required to tune any kind parameter. This in
contrast with the BSI methods recently proposed in~\cite{ohlsson2014blind,ahmed2014blind}, where, although the system is retrieved
via a convex optimization problem, the user is required to select some
regularization parameters and the model order. The method derived in this paper
follows the same approach used in~\cite{risuleo2015kernel}, where a novel
method for Hammerstein system identification is described.

The paper is organized as follows. In the next section, we introduce the BSI
problem and we state our working assumptions. In Section~\ref{sec:Bayesian}, we
give a background on kernel-based methods, while in Section~\ref{sec:marginal}
we describe our approach to BSI\@. Section~\ref{sec:experiments} presents some numerical experiments, and some conclusions end the paper.

\section{Blind system identification}\label{sec:BSI}
\begin{figure}[!ht]
\begin{center}
{\includegraphics[width=5cm]{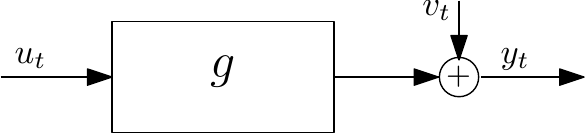}}
 \caption{\emph{Block scheme of the system identification scenario.}}\label{fig:block_scheme}
\end{center}
\end{figure}

We consider a SISO linear time-invariant discrete-time dynamic system (see
Figure~\ref{fig:block_scheme})
\begin{equation}\label{eq:sys1}
y_t = \sum_{i=0}^{+\infty} g_i u_{t-i} + v_t \,,
\end{equation}
where ${\{g_t\}}_{t=0}^{+\infty}$ is a strictly causal transfer function (i.e., $g_0 = 0$) representing the dynamics of the system, driven by the input $u_t$. The measurements of the output $y_t$ are corrupted by the process $v_t$, which is zero-mean white Gaussian noise with unknown variance $\sigma^2$.
For the sake of simplicity, we will also hereby assume that the system is at rest until $t=0$.

We assume that $N$ samples of the output measurements are collected, and denote
them by ${\{y_t\}}_{t=1}^{N}$. 
The input $u(t)$ is not directly
measurable and only some information about it is available. More specifically,
we assume we know that the input, restricted to the $N$ time instants ${\{u_t\}}_{t=0}^{N-1}$, belongs to a certain subspace of $\mathbb{R}^N$ and thus can be written as
\begin{equation}\label{eq:input sec}
u = Hx \,,
\end{equation}
where $u = \matrix{ u_0 & \cdots & u_{N-1} }^T$. In the above equation, $H \in \mathbb{R}^{N \times p}$ is a known matrix with full column rank and $x\in \mathbb{R}^{p}$, $p \leq N$, is
an unknown vector characterizing the evolution of $u(t)$. Below we report two examples of inputs
generated in this way.
\subsubsection*{Piecewise constant inputs with known switching instants}
Consider a piecewise constant input signal $u(t)$  with known switching
instants $T_1,T_2\dots T_p$, with $T_p = N$. The levels the input takes in
between the switching instants are unknown and collected in the vector $x$.
 Then, the input signal can be expressed as 
\begin{equation}
u= \matrix{u_0\\u_1\\\vdots
    \\u_{T_1-1}\\u_{T_1}\\\vdots\\u_{T_2-1}\\\vdots\\u_{T_{p-1}}\\\vdots\\u_{T_{p}}}
    = \matrix{1 &  & & \\1&  & &  \\ \vdots \\1 &  && \\  & 1 & & \\&\vdots &
    &\\ & 1 & &\\&  &\ddots &\\  &  && 1\\ &  &&\vdots\\&  &&
    1}\matrix{x_1\\x_2\\\vdots\\x_p}=Hx \,.
\end{equation}
with
\begin{equation}
H = \diag\, \{\mathbf{1}_{T_1},\,\mathbf{1}_{T_2-T_1},\,\ldots,\,\mathbf{1}_{T_p-T_{p-1}} \} \,, H \in \mathbb{R}^{N \times p}
\end{equation}
where $\mathbf{1}_{m}$ denotes a column vector of length $m$ with all entries equal to 1:

The vector $x$ needs to be estimated
from output data. Applications of BSI with
piecewise constant inputs are found in room occupancy
estimation~\cite{ebadat2013estimation} and nonintrusive appliance load
monitoring (NIALM)~\cite{hart1992nonintrusive,dong2013dynamical}.

\subsubsection*{Combination of known sinusoids}

Assume that $u$ is composed by the sum of $p$ sinusoids with unknown amplitude
and known frequencies $\omega_1,\,\ldots,\,\omega_p$. Then in this case we have
\begin{equation}
  H = \matrix{ \sin(\omega_1) & \cdots & \sin(\omega_p) \\  \vdots & & \vdots \\ \sin(N \omega_1) & \cdots &
  \sin(N \omega_p)} \,,
\end{equation}
with $\omega_1,\,\ldots,\,\omega_p$ such that $H$ is full column rank. The
vector $x$ represents the amplitude of the sinusoids. Applications of this
setting are found in blind channel estimation~\cite{ahmed2014blind}.

\subsection{Problem statement}
We state our BSI problem as the problem of obtaining an estimate of the impulse response $g_t$ for $n$ time instants,
namely ${\{g_t\}}_{t=1}^{n}$, given ${\{y_t\}}_{t=1}^{N}$ and $H$. Recall that,
by choosing $n$ sufficiently large, these samples can be used to approximate
$g_t$ with arbitrary accuracy~\cite{ljung1999system}. To achieve our goal we
will need to estimate the input $u = \matrix{ u_0 & \cdots & u_{N-1} }^T$;
hence, we might also see our problem as a \emph{blind deconvolution problem}.

\begin{rem}
The identification method we  propose in this paper can be derived also in the
continuous-time setting, using the same arguments as in~\cite{pillonetto2010new}. However,
for ease of exposition, here we focus only on the discrete-time case.
\end{rem}

In a condition of complete information, this problem can be solved by least
squares~\cite{ljung1999system} or using regularized kernel-based approaches
~\cite{pillonetto2010new},~\cite{pillonetto2011prediction},~\cite{chen2012estimation}. 
\subsection{Identifiability issues}\label{sec:identifiability}
It is well-known that BSI problems are not completely solvable (see
e.g.~\cite{abedmeraim1997blind,gustafsson1995blind}). This because the system and the input can be determined up to a scaling factor, in the sense that every pair $(\alpha u, \frac{1}{\alpha} g)$,
$\alpha \in \mathbb{R}$, can describe the output dynamics equally well. Hence,
we shall consider our BSI problem as the problem of
determining the system and the input up to a scaling factor. Another possible
way out for this issue is to assume that $\|g\|_2$ or $g_1$ are known~\cite{bai2004convergence}.

\section{Kernel-based system identification}\label{sec:Bayesian}
In this section we briefly review the kernel-based approach introduced
in~\cite{pillonetto2010new,pillonetto2011prediction} and show how to readapt it to BSI problem.

Let us first introduce the following vector notation
$$
u := \begin{bmatrix} u_0 \\ \vdots \\ u_{N-1} \end{bmatrix} \,, y := \begin{bmatrix} y_1 \\ \vdots \\ y_N \end{bmatrix} \,,\, g := \begin{bmatrix} g_1 \\ \vdots \\ g_n \end{bmatrix} ,\, v := \begin{bmatrix} v_1 \\ \vdots \\ v_N \end{bmatrix}
$$
and  the operator $\T_n(\cdot)$ that, given a vector of length $N$, maps it to an
$N\times n$ Toeplitz matrix, e.g.
$$
\T_n(u) = \begin{bmatrix} u_0 & 0  & & \cdots & 0 \\ u_1 & u_0 & 0 & \cdots & 0 \\ \vdots & \vdots &  &\ddots & \vdots \\ u_{N-2} & u_{N-3}  &  \cdots & u_{N-n+1}& 0 \\ u_{N-1} & u_{N-2}  &  \cdots &\cdots &  u_{N-n}  \end{bmatrix} \, \in \, \mathbb{R}^{N \times n}  \,.
$$
We shall reserve the symbol $U$ for $\T_n(u)$. Then, the input-output relation for the available samples can be written
\begin{equation}\label{eq:sys2}
y = Ug + v  \,.
\end{equation}

In this paper we adopt a Bayesian approach to the BSI problem. Following a
Gaussian process regression approach~\cite{rasmussen2006gaussian}, we model the impulse response as follows
\begin{equation}\label{eq:prior_g}
    g \sim \N(0,\lambda K_{\beta})
\end{equation}
where $K_\beta$ is a covariance matrix whose structure depends on a shaping
parameter $\beta$, and $\lambda \geq 0$ is a scaling factor, which regulates the
amplitude of the realizations from~\eqref{eq:prior_g}. Given the identifiability
issue described in Section~\ref{sec:identifiability}, $\lambda$ can be arbitrarily set to 1.
In the context of Gaussian regression, $K_\beta$ is usually called a \emph{kernel} and its structure is crucial in imposing properties on the realizations
drawn from~\eqref{eq:prior_g}. An effective choice of kernel for system
identification purposes is given by the so-called \emph{stable spline kernels}~\cite{pillonetto2010new,pillonetto2011prediction}. In particular, in this paper we adopt the so-called
\emph{first-order stable spline kernel} (or \emph{TC kernel}
in~\cite{chen2012estimation}), which is defined as
\begin{equation}\label{eq:ssk1}
  {\{K_\beta\}}_{i,j} := \beta^{ \max(i,j)} \,,
\end{equation}
where $\beta$ is a scalar in the interval $[0,\,1)$. 
  Such a parameter regulates the decaying velocity of the generated impulse responses. 

Recall the assumption introduced in Section~\ref{sec:BSI} on the Gaussianity of noise. Due to this assumption, the joint distribution of the vectors $y$ and $g$ is  Gaussian, provided that the vector $x$ (and hence the input $u$), the noise variance $\sigma^2$ and the parameter $\beta$ are given. Let us introduce the vector
\begin{equation}\label{eq:hyperparameters}
\theta := \begin{bmatrix} x^T  & \sigma^2 & \beta \end{bmatrix} \quad \in \mathbb{R}^{p+2} \,,
\end{equation}
which we shall call \emph{hyperparameter vector}. Then we can write
\begin{equation}\label{eq:joint_Gaussian}
p\left(\left.\begin{bmatrix} y \\ g \end{bmatrix}\right|\theta \right) \sim \mathcal N \left( \begin{bmatrix} 0\\0 \end{bmatrix} , \begin{bmatrix} \Sigma_y & \Sigma_{yg} \\ \Sigma_{gy} &  K_\beta \end{bmatrix} \right)\,,
\end{equation}
where $\Sigma_{yg} = \Sigma_{gy}^T =   U K_\beta$ and $\Sigma_y =  U K_\beta U^T + \sigma^2I$.
It follows that the posterior distribution of $g$ given $y$ (and $\theta$) is Gaussian, namely
\begin{equation}\label{eq:pg}
p(g|y,\,\theta) = \mathcal N \left(Cy,\,P \right) \,,
\end{equation}
where
\begin{equation}\label{eq:CandP}
  P = {\left( \frac{U^T U}{\sigma^2} +  K_\beta^{-1} \right)}^{-1} \quad,\quad C = P \frac{U^T}{\sigma^2}  \,.
\end{equation}
From~\eqref{eq:pg}, the impulse response estimator can be derived as the
Bayesian estimator~\cite{anderson2012optimal}
\begin{equation}\label{eq:Bayesest_gaussian}
\hat g = \mathbb E [g|y,\,\theta] = C y \,.
\end{equation}
%
%
Clearly, such an estimator is a function of $\theta$, which needs to be determined from the available data $y$ before performing the estimation of $g$. Thus, the BSI algorithm we propose in this paper consists of the following steps.
\begin{enumerate}
  \item Estimate the hyperparameter vector $\theta$.
  \item Obtain $\hat g$ by means of~\eqref{eq:Bayesest_gaussian}.
\end{enumerate}

In the next section, we discuss how to efficiently compute the first step of
the algorithm.

\section{Estimation of the hyperparameter vector}\label{sec:marginal}

An effective approach to choose the hyperparameter vector characterizing the
impulse response estimator~\eqref{eq:Bayesest_gaussian} relies on Empirical
Bayes arguments~\cite{maritz1989empirical}. More precisely, since $y$ and $g$
are jointly Gaussian, an efficient method to choose $\theta$ is given by
maximization of the marginal likelihood~\cite{pillonetto2014tuning}, which is obtained by integrating out $g$ from the joint probability density of $(y,\,g)$. Hence, an estimate of $\theta$ can be computed as
follows\begin{equation}
  \hat \theta = \arg\max_\theta \log p(y|\theta) \,.
 \label{eq:max_marginal}
\end{equation}

Solving~\eqref{eq:max_marginal} in that form can be hard, because it is a
nonlinear and non-convex problem involving a large number ($p+2$) of decision variables. For this reason, we propose an iterative solution scheme which resorts to the EM method. To this
end, we define the complete likelihood
\begin{equation}\label{eq:complete_likelihood}
L(y,\,g|\theta)  := \log p(y,\,g|\theta) \,,
\end{equation}
which depends also on the \emph{missing data} $g$.
Then, the EM method provides $\hat \theta$ by iterating the following steps:

\begin{description}
  \item[(E-step)] Given an estimate $\hat \theta^k$ after the $k$-th iteration of the scheme, compute
    \begin{equation}
      \mathcal{Q}(\theta,\,\hat\theta^k) := \E_{p(g|y,\,\hat\theta^k)}\left[L(y,\,g|\theta) \right] \,;
    \end{equation}
  \item[(M-step)] Compute
    \begin{equation}
      \hat \theta^{k+1} = \arg\max_\theta \mathcal{Q}(\theta,\,\hat\theta^k) \,.
    \end{equation}
\end{description}
The iteration of these steps is guaranteed to converge to a (local or global)
maximum of~\eqref{eq:max_marginal}~\cite{mclachlan2007algorithm}, and the iterations
can be stopped if $\|\hat\theta^{k+1} - \hat \theta^k\|_2$ is below a given
threshold.

Assume that, at  iteration $k+1$ of the EM scheme, the estimate
$\hat \theta^{k}$ of $\theta$ is available. Using the current estimate of the
hyperparameter vector, we construct the matrices $\hat C^{k}$ and $\hat P^{k}$
using~\eqref{eq:CandP} and, accordingly, we denote by $\hat g^{k}$ the estimate
of $g$ computed using~\eqref{eq:Bayesest_gaussian}, i.e. $\hat g^{k} = \hat C^{k}y$ and the linear prediction of $y$ as $\hat y^k = U\hat g^{k}$. Furthermore, let us define
\begin{align}\label{eq:matrix_A_B}
\hat A^k & = -H^T\R^T \left((\hat P^k + \hat g^{k}\hat g^{kT})\kron I_{N} \right)\R H \nonumber\\
  \hat b^k & =  H^T {\T_N(\hat g^{k})}^T y \,, 
\end{align}
where $\R\in \mathbb{R}^{Nn\times N}$ is a matrix such that, for any $u\in\mathbb{R}^N$:
\begin{equation}\label{eq:matrix_R}
  \R u =   \vect ( \T_n(u) ) \,.
\end{equation}
Having introduced this notation, we can state the following theorem, which provides a set of upgrade rules to obtain the hyperparameter vector estimate $\hat \theta^{k+1}$.

\begin{theorem}\label{th:main}
Let $\hat \theta^{k}$ be the estimate of the hyperparameter vector after the $k$-th iteration of the EM scheme. Then 
\begin{equation}
\hat \theta^{k+1} = \begin{bmatrix} \hat x^{k+1,T} & \hat \sigma^{2,k+1} & \hat \beta^{k+1} \end{bmatrix}
\end{equation} 
can be obtained performing the following operations:
\begin{itemize}
\item The input estimate is updated computing
\begin{equation}\label{eq:new_x}
  \hat x^{k+1} = -{(A^k)}^{-1}b^k \,;
\end{equation}
\item The noise variance is updated computing
\begin{equation}\label{eq:new_sigma}
\hat \sigma^{2,k+1} = \frac{1}{N} \left(\|y - \hat y^k \|_2^2 + \trace\left[\hat U^{k+1} \hat P^{k} \hat U^{k+1,T} \right]\right)\,, 
\end{equation}
where $\hat U^{k+1}$ denotes the Toeplitz matrix of the sequence $\hat u^{k+1} = H\hat x^{k+1}$;
\item The kernel shaping parameter is updated solving
\begin{equation}\label{eq:new_beta}
\hat \beta^{k+1} = \arg \min_{\beta \in [0,\,1)} 
\mathcal{Q}(\beta,\hat \theta^k)\,,
\end{equation}
where 
\begin{equation}\label{eq:Q_beta}
\!\!\!\mathcal{Q}(\beta,\hat \theta^k) =  \log\det K_\beta+\trace \left[K_\beta^{-1}(\hat P^k + \hat g^{k}\hat g^{kT})\right] .
\end{equation}
\end{itemize}
\end{theorem}
Hence, the maximization problem~\eqref{eq:max_marginal} reduces to a sequence of
very simple optimization problems. In fact, at each iteration of the EM
algorithm, the input can be estimated by computing a simple update rule
available in closed-form. The same holds for the noise variance, whereas the
update of the kernel hyperparameter $\beta$ does not admit any closed-form
expression. However, it can be retrieved by solving a very simple scalar
optimization problem, which can be solved efficiently by grid search, since the
domain of $\beta$ is the interval $[0,\,1)$. 

It remains to establish a way to set up the initial estimate $\hat \theta^0$ for
the EM method. This can be done by just randomly choosing the entries of such a
vector, keeping the constraints $\hat \beta^0 \in [0,\,1)$, 
$\hat \sigma^{2,0} >
0$.

Below, we provide our BSI algorithm.
\begin{algorithm}[ht!]\label{alg}
\textbf{Algorithm}: Bayesian kernel-based EM Blind System Identification  \vspace{0.1cm}\\
Input: ${\{y_t\}}_{t=1}^N$, $H$\vspace{0.1cm}\\
Output: ${\{\hat{g}\}}_{t=1}^n$, ${\{\hat u_t\}}_{t=1}^N$
\begin{enumerate}
  \item Initialization: randomly set $\hat \theta^{0} =
    \matrix{\hat x^{0T}& \hat \sigma^{2,0}  & \hat \beta^{0}}$
\item Repeat until convergence:
  \begin{enumerate}
    \item  \textbf{E-step:} update $\hat P^{k}$, $\hat C^{k}$
      from~\eqref{eq:CandP} and $\hat g^{k}$ from~\eqref{eq:Bayesest_gaussian};
    \item  \textbf{M-step:} update the parameters:
      \begin{itemize}
        \item $\hat x^{k+1}$ from~\eqref{eq:new_x};
        \item $\hat \sigma^{k+1}$ from~\eqref{eq:new_sigma},
        \item $\hat \beta^{k+1}$ from~\eqref{eq:new_beta}
      \end{itemize}
  \end{enumerate}
\item Compute ${\{\hat{g}\}}_{t=1}^n$ from~\eqref{eq:Bayesest_gaussian}
  and ${\{\hat u_t\}}_{t=1}^N = H \hat x$;
\end{enumerate}
\end{algorithm}

\section{Numerical experiments}\label{sec:experiments}
We test the proposed BSI method by means of Monte Carlo experiments.
Specifically, we perform 6 groups of simulations where, for each group, 100
random systems and input/output trajectories are generated. The random systems
are generated by picking 20  zeros and 20 poles with random magnitude and phase.
The zero magnitude is less than or equal to 0.95  and the pole magnitude is no
larger than 0.92. The inputs are piecewise constant signals and the number of
switching instants is $p = 10,\,20,\,30,\,40,\,50,\,60$, depending on the group
of experiments. We generate 200 input/output samples per experiment. The output
is corrupted by random noise whose variance is such that $\sigma^2 =
\var(Ug)/10$, i.e.\ the noise variance is ten times smaller than the variance of
the noiseless output. The goal of the experiments is to estimate $n=50$ samples
of the generated impulse responses. We compare the following three estimators.
\begin{itemize}
\item {\bf B-KB}. This is the proposed Bayesian kernel-based BSI method that estimates the input by marginal likelihood maximization, using an EM-based scheme. The convergence criterion of the EM method is $\|\hat{\theta}^{k+1} - \hat{\theta}^{k} \|_2 < 10^{-3}$.
\item {\bf NB-LS}. This is an impulse response estimator based on the least
  squares criterion. Here, we assume that the input is known, so the only
  quantity to be estimated is the system. Hence, this corresponds to an unbiased
  FIR estimator~\cite{ljung1999system}.
\item {\bf NB-KB}. This is the Bayesian kernel-based system identification
  method introduced in~\cite{pillonetto2010new} and revisited
  in~\cite{chen2012estimation}. Like the estimator NB-LS, this method has knowledge of the input and in fact corresponds to the estimator B-KB when $x$ is known and not estimated.
\end{itemize}
The performance of the estimators are evaluated by means of the output fitting score
\begin{equation}
FIT = 1- \frac{\|\hat{U}_i \hat{g}_i - U_i g_i\|_2}{\|U_i g_i - \overline{U_i g_i}\|_2} \,, 
\end{equation}
where, at the $i$-th Monte Carlo run, ${U}_i$ and $\hat{U}_i$ are the Toeplitz
matrices of the true and estimated inputs ($ \hat{U}_i = {U}_i$ if the method
needs not estimate $x$),  ${g}_i$ and $\hat{g}_i$ are the true and estimated
systems and $\overline{U_i g_i}$ is the mean of $U_i g_i$.

Figure 2 shows the results of the six group of experiments. As expected, the estimator NB-KB, which has access to the true input, gives the best performance for all the values of $p$ and in fact is independent of such a value. Surprisingly, for $p = 10$ amd $p = 20$, the proposed BSI method outperforms the least squares estimator that knows the true input. An example of one such Monte Carlo experiments in reported in Figure 3.
\begin{figure}[!ht]\label{fig:experiments}
\begin{center}
\includegraphics[width=0.45\textwidth]{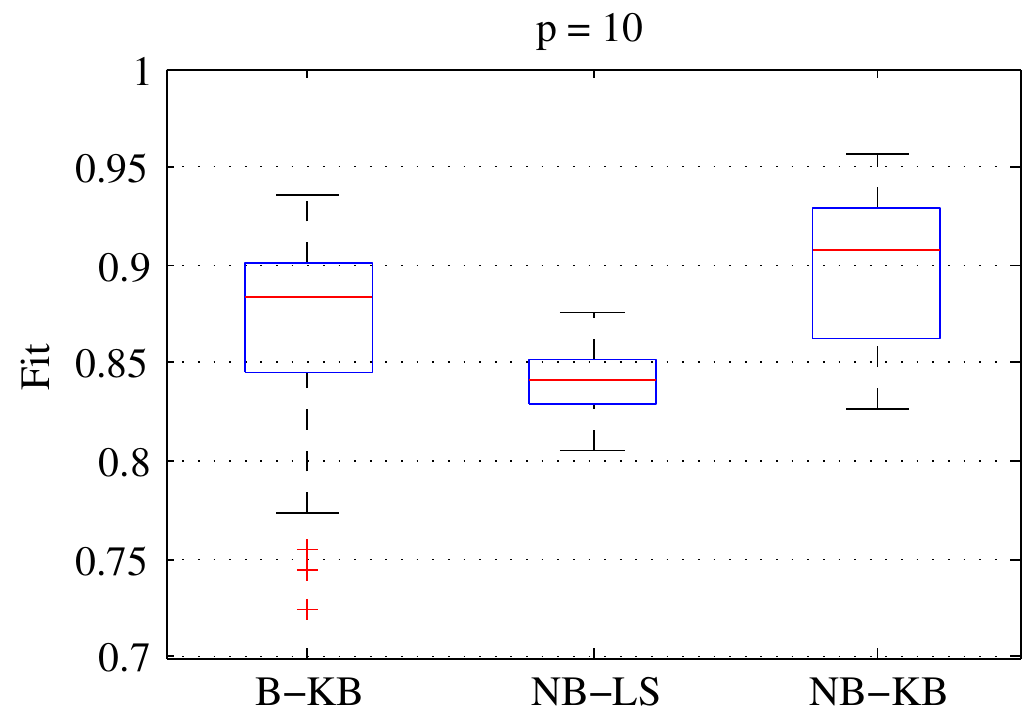} \includegraphics[width=0.45\textwidth]{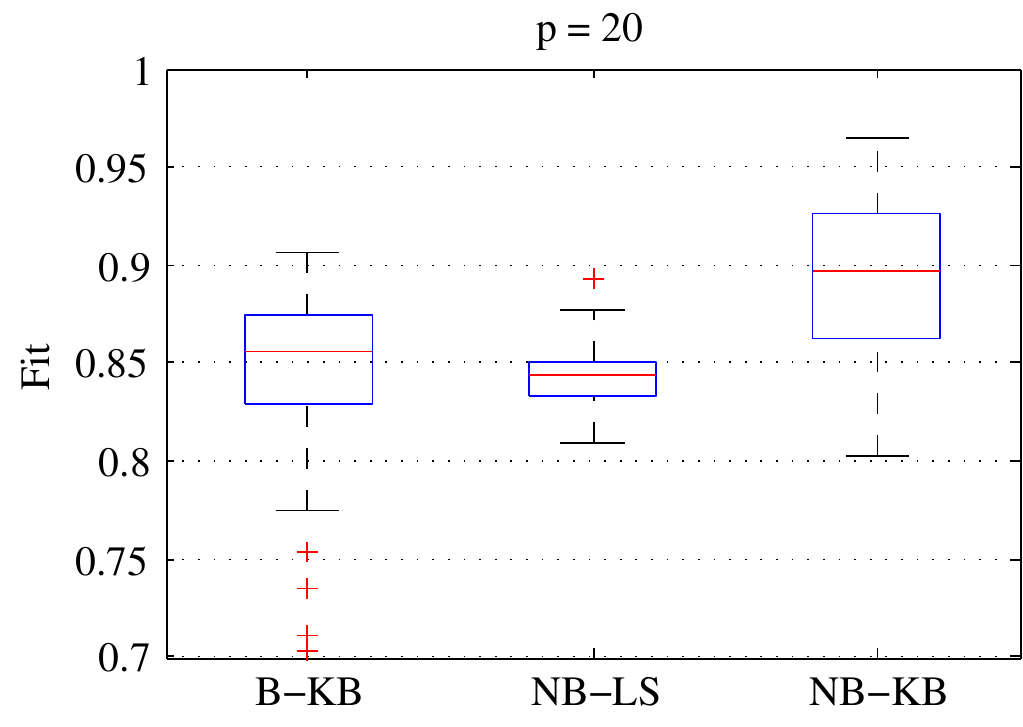} \\
  \includegraphics[width=0.45\textwidth]{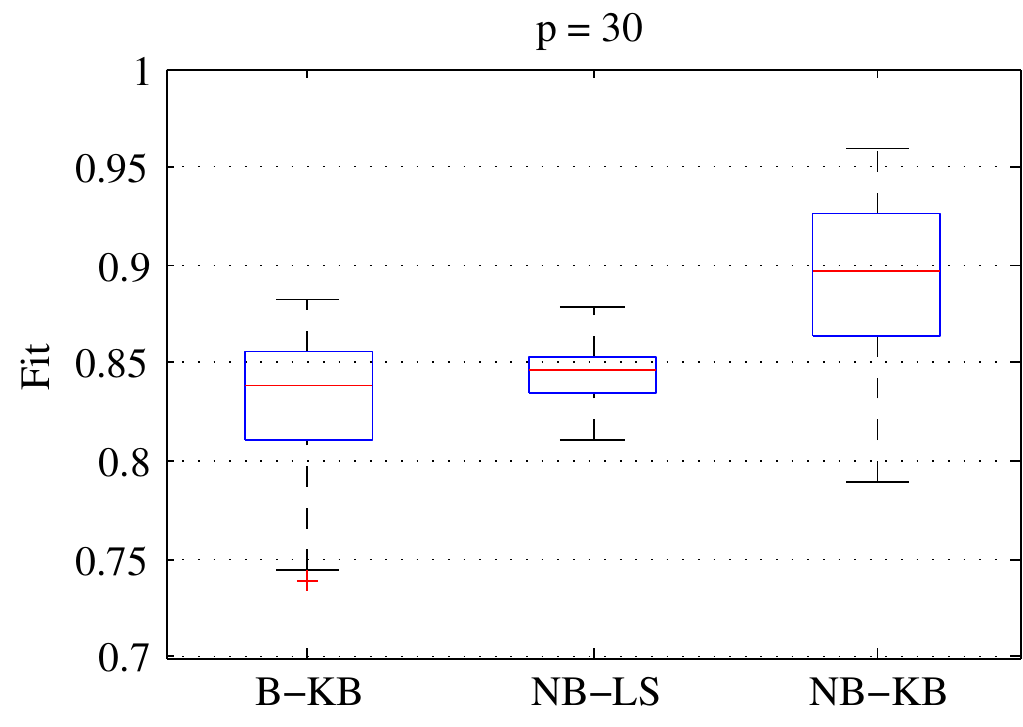}
  \includegraphics[width=0.45\textwidth]{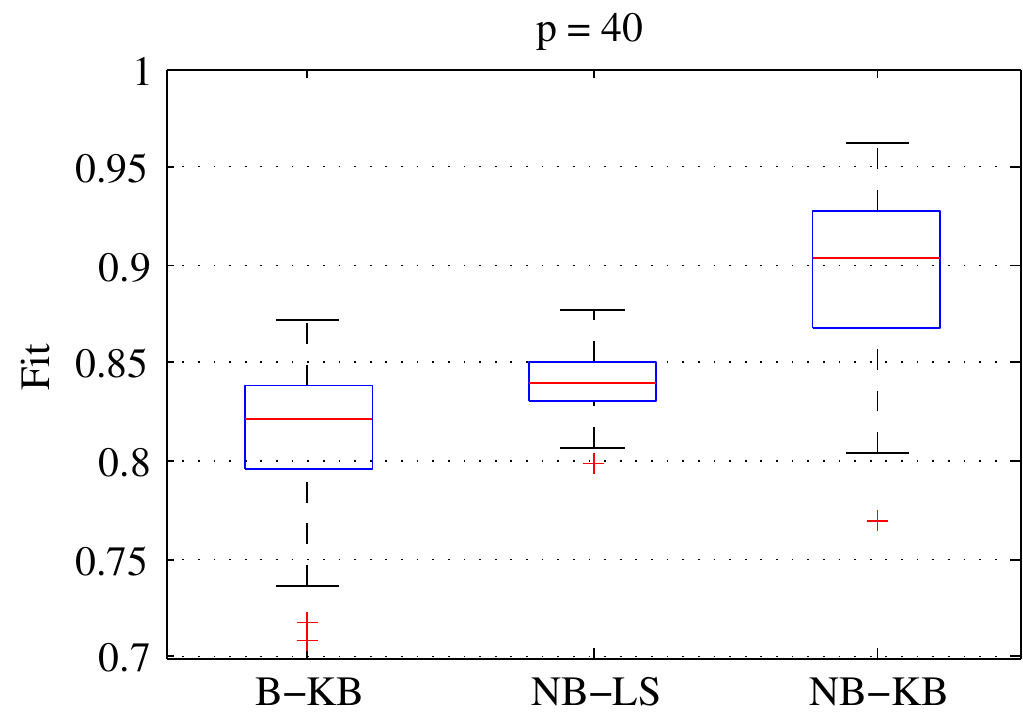} \\ \includegraphics[width=0.45\textwidth]{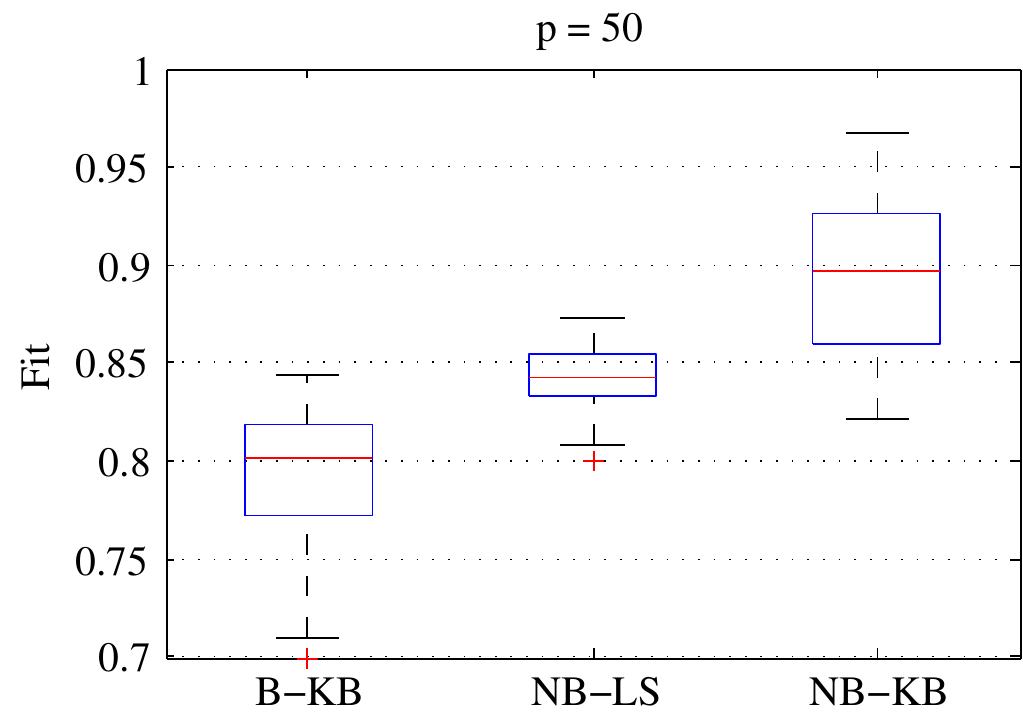} \includegraphics[width=0.45\textwidth]{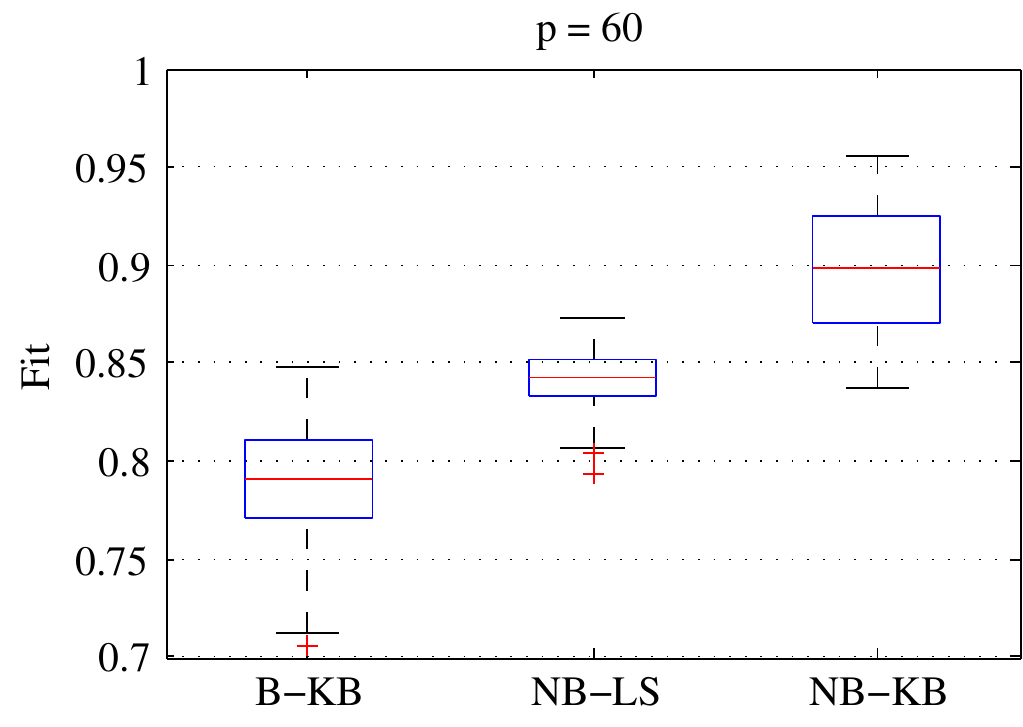}
\caption{Results of the Monte Carlo simulations for different values of $p$, namely the number of different levels in the input signals.}
\end{center} 
\end{figure}
\begin{figure}[!h]\label{fig:example}
\begin{center}
\begin{tabular}{c}
\includegraphics[width=8.5cm]{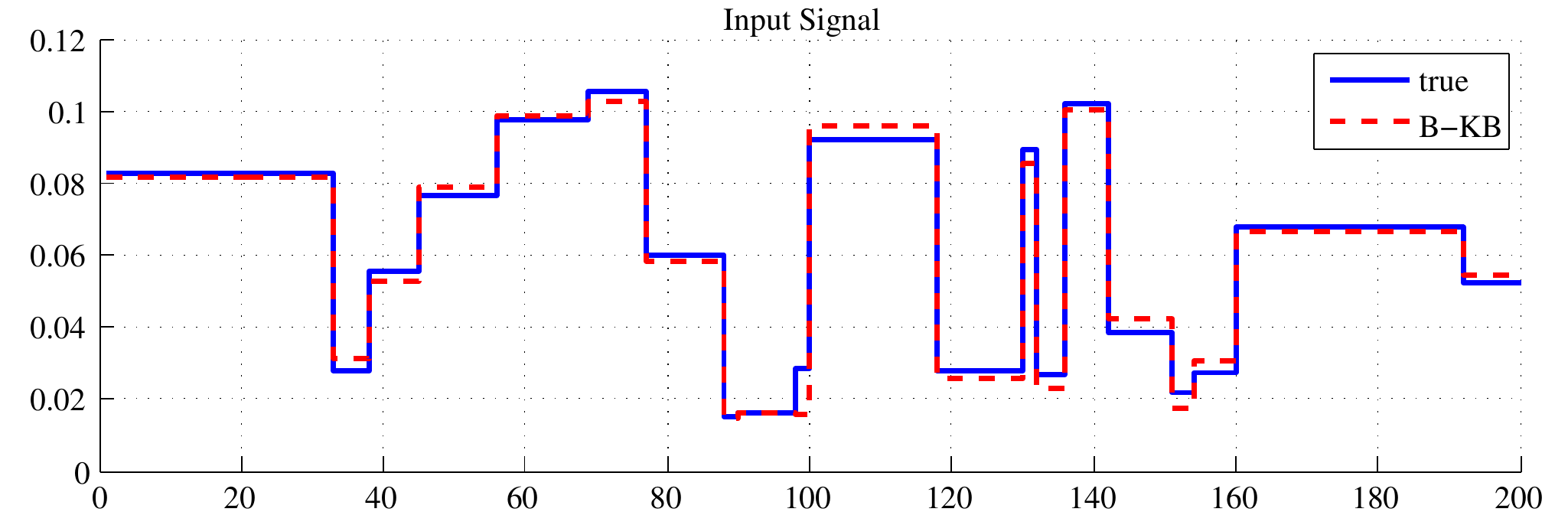} \\ \includegraphics[width=8.5cm]{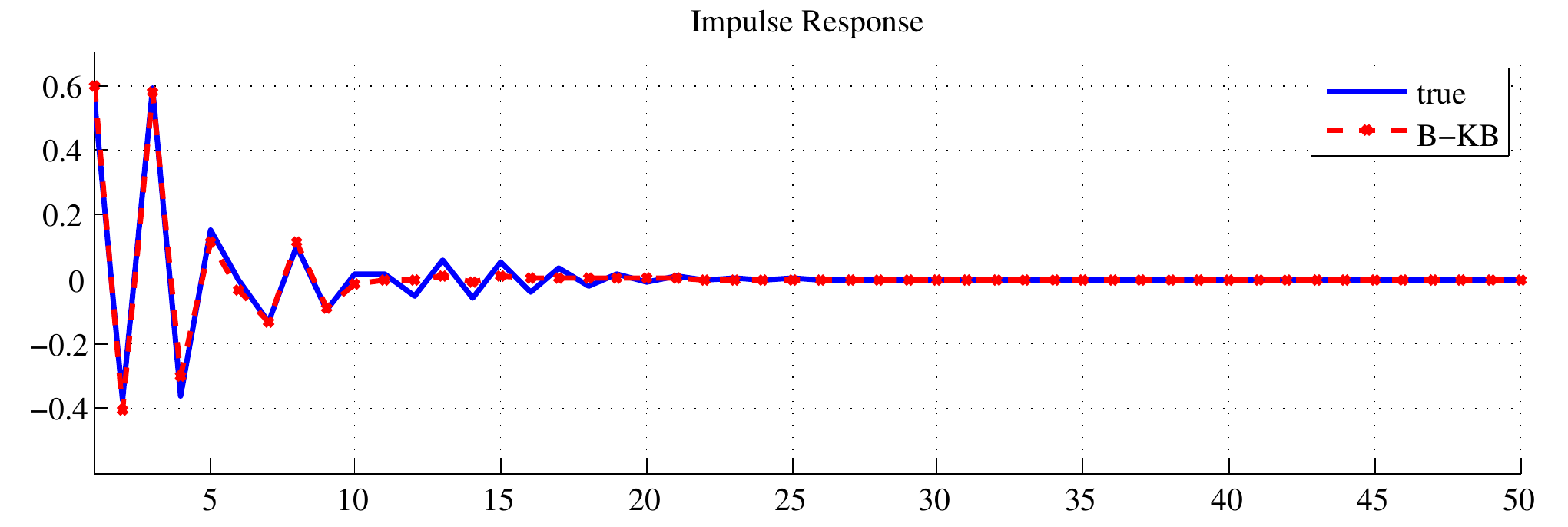} \\
\includegraphics[width=8.5cm]{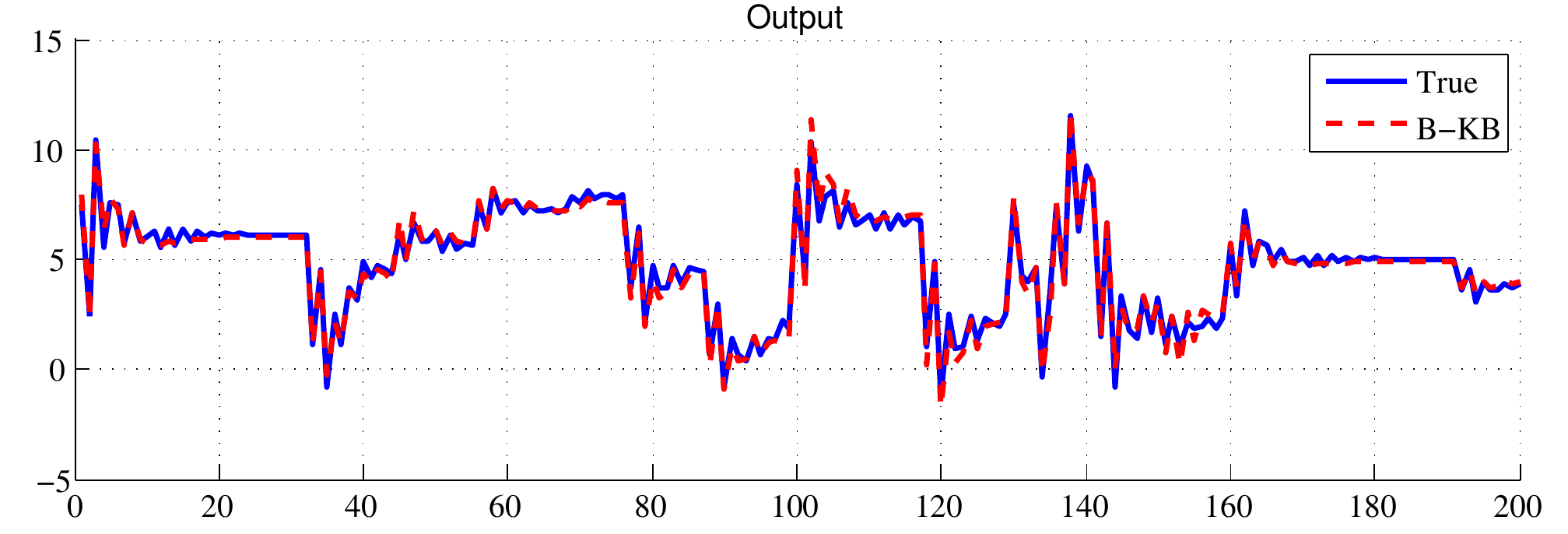} 
\end{tabular}
\caption{Example of one Monte Carlo realization with $p=20$. Top panel: true vs estimated (normalized) input. Middle panel: true vs estimated (normalized) impulse response. Bottom panel: true vs predicted output.}
\end{center} 
\end{figure}
As one might expect, there is a performance degradation as $p$ increases, since the blind estimator has to estimate more parameters. Figure 4 shows the median of the fitting score of each group of experiments as function of $p$. It appears that, approximately, there is a linear trend in the performance degradation.
\begin{figure}[!ht]\label{fig:performance_p}
\begin{center}
\includegraphics[width=6.5cm]{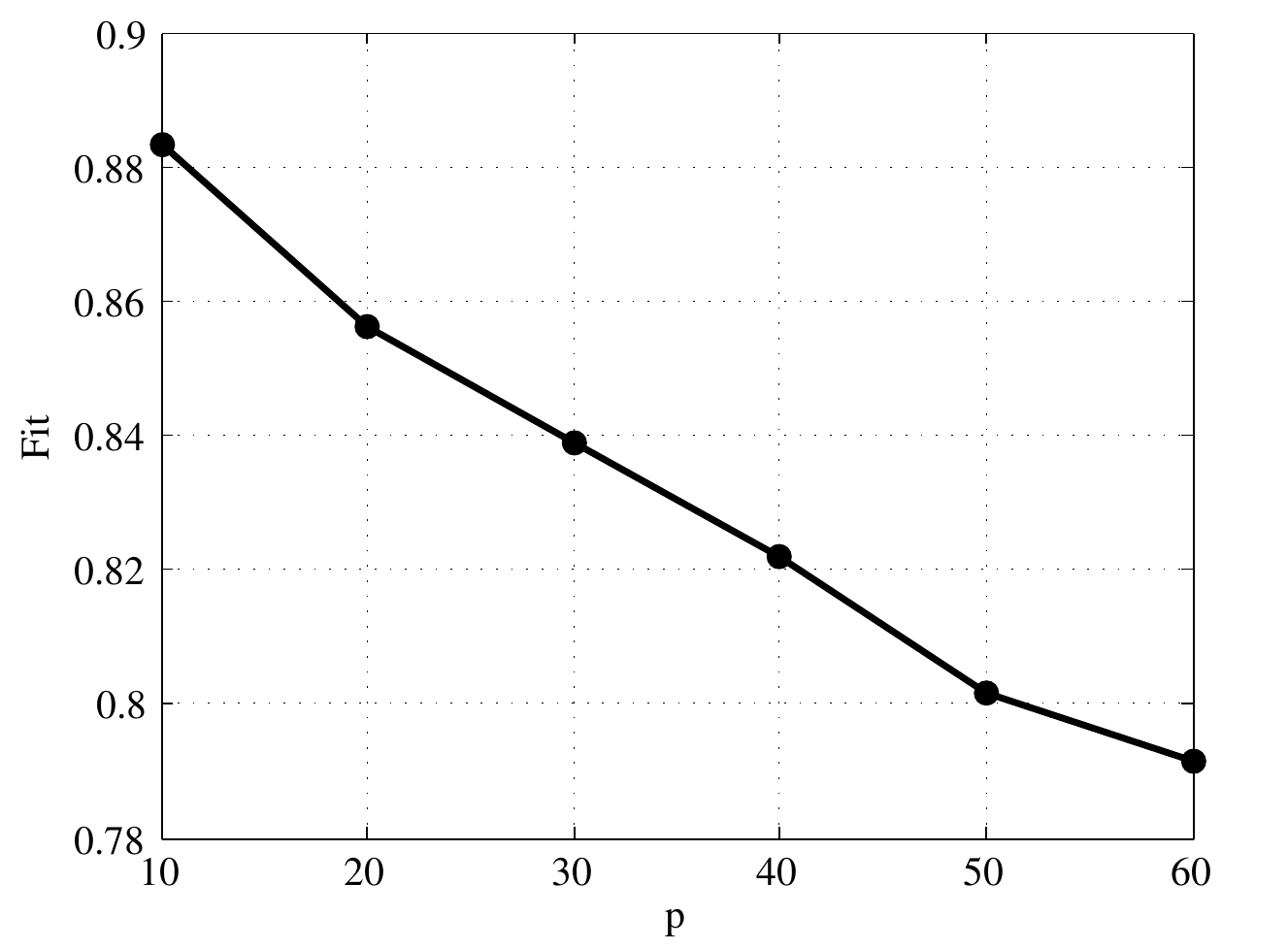}
\caption{Median of the fitting score for each group of Monte Carlo simulations as function of $p$.}
\end{center}
\end{figure}
\section{Conclusions}
In this paper we have proposed a novel blind system identification algorithm. Under a Gaussian regression framework, we have modeled the impulse response of the unknown system as the realization of a Gaussian process. The kernel chosen to model the system is the stable spline kernel. We have assumed that the unknown input belongs to a known subspace of the input space. The estimation of the input, together with the kernel hyperparameter and the noise variance, has been performed using an empirical Bayes approach. We have solved the related maximization problem resorting to the EM method, obtaining a set of update rules for the parameters which is simple and elegant, and permits a fast computation of the estimates of the system and the input. We have shown, through some numerical experiments, very promising results.

We plan to extend the current method in two ways. First, a wider class of models of the system, such as the Box-Jenkins model, will be considered. We shall also attempt to remove the assumption on the input belonging to a known subspace by adopting suitable Bayesian models.

\appendix

\section{Proof of Theorem~\ref{th:main}}
First note that $p(y,\,g|\theta) = p(y|g,\,\theta)p(g|\theta)$. Hence we can write the complete likelihood as
\begin{equation}
  L(y,\,g|\theta) = \log p(y|g,\,\theta) + \log p(g|\theta)
\end{equation}
and so
\begin{equation*}
  \begin{split}
    L(y,\,g|\theta) &=  -\frac{N}{2}\log \sigma^2 - \frac{1}{2\sigma^2}\norm{y - Ug}^2 \\
					 &\quad - \frac{1}{2}\log\det K_\beta - \frac{1}{2}g^T K_\beta^{-1}g  \\
     & = -\frac{N}{2}\log \sigma^2 - \frac{1}{2\sigma^2}\left( y^T y +  g^T U^T Ug  -2y^T Ug    \right) \\
     & \quad   - \frac{1}{2}\log\det K_\beta - \frac{1}{2}g^T K_\beta^{-1}g      \,.
  \end{split}
\end{equation*}
We now proceed by taking the expectation
of this expression with respect to the random variable $g|y,\hat \theta^k$. We obtain the following components
\begin{align*}
  (a)&: \E\left[ -\frac{N}{2}\log\sigma^2 \right] \!=\!
    -\frac{N}{2}\log\sigma^2 \allowdisplaybreaks[1] \nonumber\\
  (b)&: \E\left[ -\frac{1}{2\sigma^2}y^T y \right] \!=\!
    -\frac{1}{2\sigma^2}y^T y \allowdisplaybreaks[1] \nonumber \\
  (c)&: \E\left[ -\frac{1}{2\sigma^2}g^T U^T Ug \right] \!=\! \trace\left[
    -\frac{1}{2\sigma^2}U^T U(\hat P^k\! +\! \hat g^k\hat g^{k T}) \right] \allowdisplaybreaks[1] \nonumber\\
  (d)&: \E\left[ \frac{1}{\sigma^2}y^T Ug \right] \!=\!
    \frac{1}{\sigma^2}y^T U \hat g^k \allowdisplaybreaks[1] \nonumber\\ 
  (e)&: \E\left[ -\frac{1}{2}\log\det K_\beta \right] \!=\!
    -\frac{1}{2}\log\det K_\beta \allowdisplaybreaks[1] \nonumber\\
  (f)&: \E\left[ -\frac{1}{2}g^T K_{\beta}^{-1}g \right] =
    -\frac{1}{2}\trace \left[ K_\beta^{-1}( \hat P^k +  \hat g^k\hat g^{kT})  \right]
 \label{eq:E-step}
\end{align*}
It follows that $Q(\theta,\,\hat \theta^{k})$ is the summation of the elements obtained above.
By inspecting the structure of $Q(\theta,\,\hat \theta^{k})$, it can be seen that such a function splits in two independent terms, namely
\begin{equation}
  \mathcal{Q}(\theta,\,\hat\theta^k) = \mathcal{Q}_1(x,\,\sigma^2,\,\hat\theta^k) + \mathcal{Q}_\beta(\beta,\,\hat\theta^k) \,,
\end{equation}
where
\begin{equation}\label{eq:Q_1}
    \mathcal{Q}_1(x,\,\sigma^2,\,\hat\theta^k) = (a) +(b)+ (c)+ (d) 
\end{equation}
is function of $x$ and $\sigma^2$, while
\begin{equation}\label{eq:Q_beta_appendix}
    \mathcal{Q}_\beta(\beta,\,\hat\theta^k)  =  (e)+ (f)
\end{equation}
depends only on $\beta$ and corresponds to~\eqref{eq:Q_beta}. We now address the
optimization of~\eqref{eq:Q_1}. To this end we write
\begin{align}
  \mathcal{Q}_1(x,\sigma^2,\,\hat\theta^k) & = \frac{1}{\sigma^2}\mathcal{Q}_x(x,\,\hat\theta^k) + \mathcal{Q}_{\sigma^2}(\sigma^2,\,\hat\theta^k) \\
   & = \frac{1}{\sigma^2} \left(\trace\left[
    -\frac{1}{2}U^T U(\hat P^k\! +\! \hat g^k\hat g^{k T}) \right] + y^T U \hat g^k \right) \nonumber\\
    & \quad -\frac{N}{2}\log\sigma^2 -\frac{1}{2\sigma^2}y^T y \,. \nonumber
 \label{eq:Q_1_split}
\end{align}

This means that the optimization of $\mathcal{Q}_1$ can be carried out first
with respect to $x$, optimizing only the term $\mathcal{Q}_x$, which is
independent of $\sigma^2$ and can be written in a quadratic form
\begin{equation}\label{eq:quadratic_cost}
  \mathcal{Q}_x(x,\,\hat\theta^k) =  \frac{1}{2}x^T\hat A^k x + \hat b^{kT}x \,.
\end{equation}
To this end, first note that, for all $v_1\in \mathbb{R}^n$, $v_2 \in
\mathbb{R}^m$,
\begin{equation}
  \T_m(v_1)v_2 = \T_n(v_2)v_1 \,.
\end{equation}
Recalling~\eqref{eq:matrix_R}, we can write
\begin{align}
 & \trace \left[U^T U( \hat P^k\!\! +\!\!  \hat g^k\hat g^{kT})  \right] \!\!=
  \!\! {\vect(U)}^T\! (( \hat P^k \!\!+\!\!  \hat g^k\hat g^{kT}) \kron I_{N} ) \vect(U) \nonumber\\
    &= -\frac{1}{2}u^T\R^T\left(( \hat P^k +  \hat g^k\hat g^{kT}) \kron
      I_{N}\right)\R u\\
    &= -\frac{1}{2}x^T H^T \R^T\left(( \hat P^k +  \hat g^k\hat g^{kT}) \kron I_{N} \right)\R Hx \,,\nonumber
\end{align}
where the matrix in the middle corresponds to $\hat A^k$ defined in~\eqref{eq:matrix_A_B}. For the linear term we find
\begin{equation}
y^T U \hat g^k  =  y^T\T_N(\hat g^k)u = y^T\T_N(\hat g^k)Hx \,,
\end{equation}  
so that the term $\hat b^{kT}$ in~\eqref{eq:matrix_A_B} is retrieved
and the maximizer $\hat x^{k+1}$ is as in~\eqref{eq:new_x}.
Plugging back $\hat x^{k+1}$ into~\eqref{eq:Q_1} and maximizing with respect to
$\sigma^2$ we easily find $\hat \sigma^{2,k+1}$ corresponding to~\eqref{eq:new_sigma}. This concludes the proof.

\printbibliography{}
\end{document}